\documentclass[12pt,preprint]{aastex}

\shorttitle{The milliarcsecond-scale jet in the quasar J1625+4134}
\shortauthors{Jiang et al.}

\begin{document}

\title{The milliarcsecond-scale jet in the quasar J1625+4134}

\author{D. R. Jiang\altaffilmark{1,2},
        J. F. Zhou\altaffilmark{1,2},
        X. Y. Hong\altaffilmark{1,2},
        L. I. Gurvits\altaffilmark{3},
        Z.-Q. Shen\altaffilmark{4},
   \and Y. J. Chen\altaffilmark{1,2}
        }

\altaffiltext{1}{Shanghai Astronomical Observatory, Chinese
Academy
           of Sciences, Shanghai 200030, China}

\altaffiltext{2}{
            National Astronomical Observatories,
            Chinese Academy of Sciences, Beijing, China}

\altaffiltext{3}{
            Joint institute for VLBI in Europe, P.O. Box 2, 7990
            AA Dwingeloo, The Netherlands}

\altaffiltext{4}{
            Institute of Space and Astronautical Science, Yoshinodai
            3-1-1, Sagamihara, Kanagawa 229-8510, Japan}

\begin{abstract}
We present Very Long Baseline Array (VLBA) observations of the
radio source J1625+4134 at 22 and 15 GHz and analyze them in
concurrence with other existing VLBI data on this source. The
high resolution images at 15 and 22 GHz show a short and bending
jet which has about $270\degr$ difference in position angle with
the northern jet detected at lower frequencies. The new high
resolution data, combined with the data available in the
literature, allow us to estimate the spectral index of the
components and identify one of the compact components as the VLBI
core based on its flat spectrum between 5 and 22 GHz. Relative to
this core component, the jet appears to be bi-directional. The
proper motion measurement of the component C2 and the estimate of
the Doppler boosting factor suggest that the orientation of the
jet is close to the line of sight. The projection effect of an
intrinsically sharply bending jet within a few mas from the core
or the erratic change in the nozzle direction of the jet may
account for the uncommon bi-directional structure of the jet in
J1625+4134.
\end{abstract}


\section{INTRODUCTION}

The radio source J1625+4134 (1624+416; 4C\,41.32) was identified
as a quasar at the redshift of $z=2.55$ \citep{Pear88, Hewitt93}.
It has a $22^{m}$ optical magnitude, and has been detected at the
near infrared K band \citep {Lebo83}. The source shows a flat
spectrum between 178 MHz and 37 GHz with the spectral index
$\alpha = -0.34$ ($S_\nu\propto\nu^{\alpha}$). At the VLA angular
scale, the source extends to about $1\arcsec$ in the northern
direction with a very weak polarization indicative of the
magnetic field parallel to the jet structure \citep{Per82,
ODea88}.

The quasar J1625+4134 has been observed with VLBI at various
frequencies: 1.67 GHz \citep{Pol95}, 2.32 and 8.55 GHz
\citep{Fey97}; 5 GHz \citep{Pear88, Foma2000, Lis2001a}. On the
VLBI scale, its structure is rather uncommon: a short strong jet
towards the southwest and a long but weak jet emission extended
to the north--northwest.  The VLBI images at 1.67 and 2.32 GHz are
similar, the jet of J1625+4134 extends to the north to over 30
mas along the same position angle PA~=~$350 \degr$ as that in the
VLA images. Model fitting of the VLBI data revealed a component
in the southwest region which was not clearly separated from the
central compact emission in the images probably due to the
limited angular resolution \citep{Pol95, Fey97}. The VLBI images
at 5 and 8.55 GHz showed a compact core, a short curved jet in
the southwest region and a weak emission to the north. The high
resolution VSOP image at 5 GHz showed the structure of the
southwest jet only, while the image made with the ground-only
array (VLBA) of the same VSOP observation  exhibited the
two-sided jet structure \citep{Lis2001a} extending both in
southwest and north directions. This kind of sharp difference
between VLBI structures at low frequencies (and also lower
angular resolutions) and higher frequencies (and higher angular
resolutions) is relatively rare in AGNs. Another similar case has
been found in the high redshift quasar 1351-017 ($z=3.707$, Frey
et al. 2002).

There are several possible explanations for this morphological
difference seen in the jet structure of J1625+4134 at different
frequencies and angular scales. It has been found that
milliarcsecond-scale jets in many radio loud sources is
misaligned with respect to the structure at the arcsecond scale
\citep{Pear88, Hong98, Cao2000}. The large $\Delta$PA (the
difference between the jet position angles measured on mas and
arcsecond scales) may be due to the projection of a bending jet or
a helical jet \citep{Conway93}. If the brightest component in the
images is the VLBI jet base, J1625+4134 can be classified as a
quasar with a sharp bend. For the jet moving away from the core
toward the northern emission, $\Delta $PA is about $90\degr$
\citep{Lis2001b} or $270\degr$ depending on the jet bending
direction.

Alternatively, the brightest component in the 5 and 8.5 GHz images
could be not the core of J1625+4134. Rather, the core could be
weak and located, e.g., at the end of the southwest emission. In
this case the brightest emission in the 5 and 8.5 GHz images may
be caused by interaction of the jet with the ambient medium. The
jet turns for about $90\degr$ to the north in this region of
strong jet-medium interaction. Such the interaction will produce a
strong shock, which further enhances the emission and flattens the
spectrum.

The third possibility is that the source has a twin-jet structure
with the radio jet lying close to the sky plane. In this case, the
relativistic boosting effect is not important, the jet and
counter-jet are both detectable. The jet emission should be more
symmetric at the higher frequencies due to the reduced free-free
absorption of the obscuring accretion disk or torus.

In this paper we present and discuss the results of VLBA
observations at 22 and 15 GHz. In Section 2, we describe the
observations and data reduction. The results are presented in
Section 3. In Section 4, we discuss the jet morphology.
Throughout this paper we adopt the Hubble constant $H_{0}$ = 65
km\, s$^{-1}$\,Mpc$^{-1}$ and the deceleration parameter
$q_{0}=0.5$, assuming the cosmological constant $\Lambda = 0$.
With these parameters, 1 milliarcsecond (mas) corresponds to
5.91~pc at the redshift of $z = 2.55$.

\section {OBSERVATIONS AND DATA REDUCTION}

The observations of J1625+4134 at 22 GHz were conducted with the
VLBA on 1 March 2000, the total on-source time was about 3 hours.
The LCP signals were recorded in 4 IF-bands for a total bandwidth
of 32 MHz with 2-bit sampling. Two strong calibrators, J1635+3805
and J1640+3946, were observed too. The data were then correlated
at the VLBA correlator in Socorro (NM, USA) and calibrated and
fringe-fitted using the NRAO Astronomical Image Processing System
(AIPS) software. Initial amplitude calibration was carried out
with the system temperature measurements and the NRAO-supplied
gain curves, before the fring-fitting, the data are corrected for
2-bit sampling errors and atmospheric opacity. The imaging and
self-calibration were carried out in DIFMAP package
\citep{Shep94}. The constant gain corrections of the stations
during the self-calibration are well in agreement with those for
the two calibrators.

The source was also observed as part of the VLBA survey at 15 GHz
in support to the VSOP Survey Programme \citep{Gur2002}. The
observation was conducted on 2 January 1999 with a total
bandwidth of 64 MHz using 8 IF-bands and 1-bit sampling; the
total on-source time was about 35 minutes (7 scans $ \times$ 5
minutes). The data on J1625+4134 were calibrated and
fringe-fitted alongside with other survey sources, but for the
purpose of this work the image was produced independently using
AIPS.

\section {RESULTS}

\subsection {Images}

The left panel in Fig.~1 presents the naturally weighted image of
J1625+4134 at 22~GHz. The estimated thermal noise ($1\sigma$) is
about 0.3 mJy/beam for 3 hours on-source integration, the lowest
contour in the image is 0.9~mJy/beam. The core-jet morphology is
consistent with those observed at 5 and 8.5~GHz, the jet bends
smoothly to the southwest from the brightest spot and the
emission becomes diffuse at about 3 mas.

The naturally weighted image at 15 GHz is shown in the right
panel of Fig.~1. The estimated thermal noise ($1\sigma$) is about
0.4~mJy/beam for 35 minutes on-source integration, the lowest
contour in the image is 1.0~mJy/beam. The general source
structure at this frequency is similar to that at 22 GHz.
However, one can notice traces of a weak emission at 15 GHz
north-northwest at a distance of about 3 mas from the brightest
component coinciding with the northern components seen in the
images at 5 and 8.5~GHz \citep{Pear88, Lis2001a, Fey97}.

The 5-component model represents well the core-jet structure at
22 GHz and is listed in Table~1. The calibrated $u,v$-data at
15~GHz were also fitted by 5 Gaussian components (see Table~1),
but there is a very low surface brightness emission in the
northern region in the residual map.  The estimated flux density
of this low brightness emission is about 14 mJy. In order to
compare our results with other observations, we follow the
labeling convention of the jet components introduced by Lister et
al. (2001a) in their 5 GHz VSOP image. The component from our
model called C2B and present in both 15 and 22~GHz data (see
Fig.~1 and Table~1) is absent in the VSOP image at 5~GHz by
Lister et al. (2001a). In Table~1, we also give the error
estimations of the flux density and the position of the model
components. The flux density errors of the model components
include the calibration uncertainty as well as the spread in
model-fitting results obtained in AIPS and DIFMAP. The error
estimation of the component positions are determined by the
differences between different model-fitting procedures and the
uncertainties estimated by the formula introduced by Fomalont
\cite{Foma88}.

The component D is the strongest and the most compact at both 15
and 22~GHz. It has a flat spectrum at the higher observing
frequencies, which, following the existing convention, allows us
to identify this component as a 'core'. At the brightness level of
1~mJy/beam, no evidence of the counter-jet emission was found in
the opposite direction of the bright jet. The weak emission in
the north-northwest region at 15 GHz does not seem to resemble a
collimated counter-jet, although such an interpretation cannot be
ruled out at this stage.

\subsection {The distribution of the VLBI component positions}

The right panel in Fig.~2 presents a plot of the position of the
jet components measured at different frequencies in the quasar
J1625+4134 relative to the core component D. The two-sided jet
structure consists of a short bright jet bending from the west to
the southwest at a few mas from the core and a long weak jet
extending to the north for more than 25~mas. If the curved
southwest jet continuously bends to the north, the overall
bending is at least $\Delta {\rm PA} \approx 270\degr$. Such an
extreme bending on the VLBI scale is relatively rare in the known
radio structures in AGNs.

The left panel of Fig.~2 shows the enlarged southwest jet. Within the
inner 2 mas from the core, the jet components at 8, 15 and 22 GHz
follow the same trajectory moving away from the core. But there is
about $15\degr$ difference in jet axis position angle between the
observations at these higher frequencies and the VSOP result at 5~GHz
\citep{Lis2001a}.

A careful inspection of our model-fitting results by allowing the
change in PA shows that the uncertainties in the determination of
the position angle for the inner 2 mas components are less than a
few degrees. We also note that our model fitting results at 15 and
22~GHz are in good agreement with those at 8~GHz \citep{Fey97}.
All this indicates that the offset of the jet axis at 5~GHz and
higher frequencies is real.

Frequency-dependent offsets in the position of jet axis at the
VLBI scale have been seen in 3C~454.3 at frequencies 15 and 86~GHz
\citep{Kric99} and could be caused by the opacity effects
enhanced in the regions of stronger jet bending. If the spectral
index of the emission region in the transversal direction of the
jet is different, the position of the fitted component may shift
transversally with the observing frequency. Alternatively, a
frequency-dependent core position offset may explain this
difference \citep{Loba98}. We also note that the restoring beam of
the VSOP observation is 1.08~mas~$\times$~0.22~mas at ${\rm PA} =
20\degr$, and the major axis of the beam is oriented along the
offset direction of the jet axis. This alignment could introduce
the apparent offset.

In addition to the 22 and 15~GHz VLBA data sets described above,
we have got an access to the 5~GHz VLBA data from the VLBApls
observations \citep{Foma2000}. The image of J1625+4134 from
VLBApls was originally model-fitted with two components
\citep{Foma2000}. In order to compare the models at 5~GHz and
higher frequencies, we fitted the self-calibrated VLBApls
$u,v$-data with a 5-component model which includes the northern
component B1 detected in both images at 2.3 and 8.5~GHz
\citep{Fey97}. The new model-fitting results are given in
Table~1. Fig.~3 shows the new model-fitting image based on the
VLBIpls data which is consistent with the image published by
Fomalont et al. \citep{Foma2000}. The positions of the five model
components for the VLBApls 5~GHz data are also shown in Fig.~2.
Since the resolution at 5~GHz is lower than that at high
frequencies and the structure of the original image at low
resolution is simple, the reduced $\chi^2$ of this model-fitting
result (for 30 seconds averaged data) is relatively high, 2.29.
The new model result cannot be considered as unique, it represents
only one of many possible fits.

\subsection {Variations of VLBI structure}

The C2 component is present in the VLBI models at 5 (both VSOP and
VLBApls observations), 8, 15 and 22~GHz. It is located at the same
position angle (about $-110\degr$) in all observations except the
VSOP one. In the latter, the component C2 is some $15\degr$ off
the position in all other data sets. If the C2 component in all
the observations indeed corresponds to the same physical
component, its apparent proper motion is $\mu$ = 0.11$\pm$0.04
mas/yr, corresponding to a $(7.4\pm2.7\,)c$ apparent transverse
velocity. The position of component C2 as a function of time and
the best fit of its apparent proper motion is shown in Fig.~4. The
estimate of proper motion should be treated with caution due to
possible opacity effects and spectral variations across the
emission region.

\subsection {Spectra of the components}

VLBI data on the source J1625+4134 mentioned above have been
obtained at different epochs (2.3 and 8.5~GHz -- at 1995.77;
5~GHz -- 1996.42 and 1998.1; 15~GHz -- 1999.0; 22~GHz --
2000.16). The UMRAO monitoring program at 4.8, 8.0 and 14.5~GHz
shows that the total flux density of the source varied for up to
30\% during the period 1995--2000. In spite of the source's
variability and non-simultaneous VLBI data at different
frequencies we attempted to compose and analyze the spectra of
the components. The spectra are shown in Fig.~5. Although these
spectra are no more than very rough estimates, they offer some
useful insight into the radio emission from the jet components.

The component D has a flat spectrum with $\alpha = 0.15\pm0.26$
between 5 and 22~GHz. The higher flux density at 2.3~GHz may be
due to the lower resolution which may allow for a contribution
from low brightness extended emission surrounding the source.
Based on its spectrum, the component D is the most likely
candidate for the 'core' of the source.

The components C2, C2B and B1 have steep spectra. The spectral
index of the component C between 5 and 22 GHz is
$\alpha=-1.3\pm0.3$ (the VSOP data is not included in this
spectral index estimate). The component C2B has $\alpha=
-1.2\pm0.3$ between 8 and 22~GHz, its turnover frequency is
probably between 5 and 8~GHz. The northern component B1 shows
similar spectrum with the component C2B, $\alpha = -1.1\pm0.3$
between 2 and 15~GHz.

To obtain a two-frequency non-simultaneous spectral index map of
the area around the core of J1625+4134, we produced the images at
15 and 22~GHz using the same $u,v$-ranges, cell size and
restoring beams. The problem of producing a spectral index map is
in determining the reference position in two images. We have
decided to use the position of peak emission of the core
component D to align the two images. An eastward shift of 0.04 mas
(0.6 pixels) in right ascension was needed in the image at 22 GHz
to align its D component with its counterpart at 15~GHz. We note
that due to the possible self-absorption in the core region, the
peaks of emission may lie in different parts of the core at
different frequencies. However, the frequency difference between
15 and 22~GHz is relatively small. If the absolute core position
is $r\propto\nu^{-1}$ \citep{Loba98}, the estimated
frequency-dependent position difference of the core in J1625+4134
is $< 0.1$~mas. Fig.~6 shows the spectral index distribution
superimposed with the contours of the 15~GHz image.

As one would expect, in spite of an unknown impact of the source
variability, the core area of the source has a flat spectrum. The jet
region has mostly steeper spectrum, with two noticeable exceptions
associated with the component C2 and the area between the components
C2B and C1. The flatter spectrum around component C2 is likely to
reflect the motion of the component outwards: the leading edge of the
component is brighter at higher frequencies resulting in the flatter
overall component spectrum. The second region of a flatter spectrum
between the components C2B and C1 might be related to the structure
variation or to the true spectral index distribution as well as the
structural change as in the case of the component C2.

\subsection {The brightness temperature of the mas-scale jet
components}

The brightness temperature $T_{b}$ of an elliptic Gaussian
component in the source rest frame is given by \citep{Shen97}

\begin{equation}
T_{b} =1.22\times 10^{12}(1+z)\frac{{S_\nu}}{{\nu_{ob}^{2} ab}},
\end{equation}

\noindent where $S_\nu$ is the flux density of the component in
Jy at the observing frequency $\nu$ in GHz, $a$ and $b$ are the
major and minor axes in mas respectively, and $z$ is the source
redshift. The brightness temperature distribution of the VLBI jet
components, measured at the different frequencies, is shown in
Fig.~7. Due to the resolution limit, the brightness temperature
of the core may be only a lower limit and the source variation
also may introduce some uncertainties in the estimate of the
brightness temperatures.

The brightness temperature of the core is plotted in both panels
of Fig.~7.  The high brightness temperature of the component D
also supports its identification as the core. The brightness
temperature decreases sharply within about 4~mas from the core in
the southwest region. However, the brightness temperature in the
northern jet is relatively constant from 4 to 25~mas. The
decrease of the brightness temperature in the inner jet may be
due to radiation losses and bending of the jet. If the brightness
temperature variations are dominated by the variation of the
viewing angle, the jet bending reduces the Doppler boosting
effect in the southwest jet, while the orientation of the jet
remains relatively unchanged in the northern region.

\section {Discussion}

\subsection {Spectral energy distribution of J1625+4134}

Using the data available in NED, we estimated the source
rest-frame spectral energy distribution (SED). An upper limit
flux density of 0.85~$\mu$Jy at 1~keV was measured by HEAO-A1
\citep{Bie87}. In the ROSAT-FIRST correlation study, Brinkmann et
al.(2000) found 5 radio sources within the resolution of ROSAT
around J1625+4134. Of these five, J1625+4134 is the nearest one
to the ROSAT pointing center (the position offset is $5\arcsec$) .
The X-ray flux measured by ROSAT from this area is $0.14\pm0.05
\times 10^{-12} \,{\rm erg\,s}^{-1}\,{\rm cm}^{-2}$. The
information on the photon index for this source is unavailable,
thus we have used the mean value of 2.46 for the quasars
\citep{Brink2000}. This leads to the estimated flux density of
0.012 $\mu$Jy at 1~keV.

The peak frequency of the synchrotron emission energy, $\nu_{\rm
peak} = 3.4\times 10^{12}$~Hz, is located in the sub-mm band, the
spectral energy at the peak frequency, $\nu L_{\nu} = 8.8\times
10^{38}$~W, is typical for the radio-loud quasars \citep{Sam96}.
The X-ray emission could not fit the synchrotron emission, most
likely this is to be attributed to the Self-Synchrotron Compton
scattering (SSC).

\subsection {Doppler boosting}

J1625+4134 is a typical radio loud quasar which shows all the
indications on the relativistic bulk motion of the plasma in its
jet, with its syncrotron emission Doppler boosted. There are
several methods of estimating the Doppler boosting factor.

Assuming that the X-ray emission mechanism is SSC scattering, one
can adopt the equation (1) of Ghisellini et al. \citep{Ghis93} to
estimate the Doppler boosting factor $\delta_{sp}$ in a uniform
sphere of the emission region. Readhead \citep{Read94} suggested
another method of estimating the Doppler boosting factor, the so
called equipartition Doppler boosting factor $\delta_{eq}$ . The
method is based on the assumption on the energy equipartition
between the radiating particles and the magnetic field.

Both methods above require VLBI core data at the turnover
frequency. It proved to be difficult since the core has a very
flat spectrum (see Fig.~5). In Table~2, we list the Doppler
boosting factor estimates using both methods by assuming VLBI
observing frequencies as the VLBI core turnover frequency. To
avoid a fitted zero axial ratio for the 8.5 GHz core
\citep{Fey97}, we used a value of 0.2 for this ratio in the
calculation.

In the equipartition case, a correction of the core size,
$\theta_{\rm d}=1.8 \sqrt{ab}$ was applied \citep{Mar87, Gui96},
while in the $\delta_{\rm sp}$ estimate, a factor of 0.8 was
adopted \citep{Gui96}. The new X-ray flux density was used in the
$\delta_{\rm sp}$ estimate and the spectral index $\alpha = -
0.75$ for the optically thin synchrotron emission was assumed.
Table~2 gives the estimates of the Doppler boosting factor for
VLBI core data obtained at different frequencies.

Jiang et al. \citep{Jiang98} investigated the inhomogeneous jet
parameters in the K\"onigl's model \citep{Koni81}. In this model,
a VLBI core is believed to be a base of the optically thick
synchrotron emission region of an inhomogenous jet. Assuming that
the X-ray emission mechanism is the SSC scattering and the
distributions of the magnetic field and the relativistic electron
number density in the jet follow the power law ($m = 1$ for the
magnetic field and $n = 2$ for the electron number density), and
using the value of the proper motion of $\mu = 0.11$~mas/yr, we
can obtain the Doppler boosting factor and estimate other
parameters of the model. The Doppler boosting factors
$\delta_{\rm jet}$ for the inhomogeneous jet model are also given
in Table~2.

These estimates of the Doppler boosting factor values cover a wide
range of VLBI core parameters obtained at different frequencies.
We assume the mean value $\delta = 5.2\pm 2.5$ as the best
estimate of the Doppler boosting factor in the jet of J1625+4134.
Combined with the proper motion mentioned above, the viewing
angle to the line of sight could be estimated as $\theta =
10\degr \pm 4\degr$, and the Lorentz factor of the jet $\gamma = 8
\pm 4$. These parameters are consistent with those for radio-loud
quasars.

\subsection {Possible explanation for the jet structure}

The proper motion, non-detection of the counter-jet and the
estimated value of the Doppler boosting factor suggest that the
orientation of the inner south-west jet is close to the line of
sight. These arguments rule out a two-sided jet morphology as an
explanation of the observed bi-directional (southwest and
north-northwest) jet structure in J1625+4134.

In radio-loud AGNs, large apparent bending angles of jet can be
explained by the projection effect, since the viewing angle to the line
of sight is small. The observed difference in the position angle of
different areas along the jet, $\Delta PA$, is related to the true
bending angle of the jet $\Delta \phi$ as

\begin{equation}
cos{\Delta\phi}=cos\theta_{1}cos\theta_{2}+sin\theta_{1}sin\theta_{2}cos\Delta
{PA},
\end{equation}

\noindent where $\theta_{1}$ and $\theta_{2}$ are the viewing
angles of the jet. The equation (2) holds only for the step
change of the $\Delta \phi$. For a curved jet, the accumulated
true bending angle $\Delta \phi$ should be integrated along the
trajectory of the bending using the small angle approximation. If
the jet moves along a cone with the half-opening angle
$\theta_{1}$ (this means that $\theta_{2}$ is equal to
$\theta_{1}$), the integrated true bending angle $\Delta \phi =
sin\theta_{1}(\Delta PA)$. In the J1625+4134 case, the estimated
viewing angle of the southwest inner jet $\theta_{1}$ is about
$10\degr$. If the two-sided jet is connected via an invisible
curved trajectory, $\Delta PA = 270 \degr$ corresponds to $\Delta
\phi = 47 \degr$. This is a large intrinsic bending.

The cause for such a large bending of the jet in J1625+4134 is not
clear. There is no obvious evidence of the enhanced emission at
the region of sharp bending region of its jet. It seems that the
bending of the jet is not caused by the interaction with the
surrounding medium, since such interaction will produce strong
shocks, further enhancing the radio emission.

Another possible explanation for the bi-directional jet structure
is the intrinsic change in the jet ejection angle caused by a
ballistic precession or erratic change in the nozzle direction.
The bi-directional jet structure in J1625+4134 is more like the
erratic change in the nozzle direction, because two jets follow
two very different ballistic trajectories \citep{Gome99}. The
northern jet may have been ejected when the nozzle of the jet
pointed to the north and the southwest jet is the result of a
newer ejection. In this case, the change of the nozzle direction
would occur more than about 50 years ago -- the time needed for
the bending southwest jet material to reach its present extension
of about 4-5~mas with the apparent proper motion of the component
C2 of about 0.1~mas/yr.

Since the northern emission has a steep spectrum, a high
resolution observations with a higher sensitivity at low
frequencies may be useful to distinguish between the two possible
models. A detection of the connection between the two jets will
support the projection effect explanation.

\section{SUMMARY}

The quasar J1625+4134 shows an uncommon bi-directional jet
structure on the VLBI scale at various observing frequencies. We
presented the images of VLBA observations at 15 and 22~GHz,
showing a short curved southwest jet which extends for several
mas from the core. Our VLBI results, combined with the data
available in the literature, allow us to estimate the spectral
index of the components and to identify the component D as the
VLBI core based on its flat spectrum between 5 and 22~GHz. The
jet components distribution at different observing frequencies
favors a large difference of about $270 \degr$ in the position
angles of the direction of the jet propagation in the southwest
and northern areas from the core. A proper motion of the
component C2 in the southwest jet, $\mu = 0.11$~mas/yr, was
estimated. Using the multi-band VLBI data on the component D and
different physical methods, we obtained the best estimate of the
Doppler boosting factor, $\delta =  5.2\pm 2.5$, corresponding to
the southwest inner jet viewing angle of $10\degr\pm 4\degr$ to
the line of sight with the Lorentz factor $\gamma = 8\pm 4$.

The proper motion, the estimated Doppler boosting factor and
non-detection of the counter-jet emission at the higher VLBI
frequencies rule out the two-sided-jet explanation for the
observed bi-directional jet morphology. According to the analysis
on the relativistic jet parameter, we point out on the
possibility that the southwest jet goes to the north via a
helical trajectory, while an erratic change in the nozzle
direction also can explain the bi-directional jet structure. High
resolution VLBI observations at lower frequencies may provide
clue on this model.

We found a difference of about $15\degr$ in the southwest jet axis
between the 5~GHz VSOP image and other VLBA observations at higher
frequencies. Our re-analysis of the 5~GHz VLBApls data narrows the
difference in the axis position, but the hypothesis on a frequency
dependent position of the jet axis needs further verification.

\begin{acknowledgements}
Support from the Chinese fund NKBRSF (No. G1999075403) is
gratefully acknowledged. The VLBA is operated by the National
Radio Astronomy Observatory which is managed by Associated
Universities, Inc., under cooperative agreement with the National
Science Foundation. This paper has made use of the NASA/IPAC
Extragalactic Database (NED), which is operated by the Jet
Propulsion Laboratory, California Institute of Technology, under
contract with the National Aeronautics and Space Administration.
We have also used the data from the University of Michigan Radio
Astronomy Observatory which is supported by the National Science
Foundation and by funds from the University of Michigan. We are
grateful to the teams of VLBApls \citep{Foma2000} and VLBA 2~cm
VSOP support survey \citep{Gur2002} for the permission to
re-analyze their data. The authors acknowledge support from the
exchange programme in radio astronomy of the Chinese and Dutch
Academies of Sciences. We wish also to thank the anonymous
referee for the valuable comments.
\end{acknowledgements}

\clearpage

\begin{deluxetable}{lccccccc}
\tabletypesize{\scriptsize}
\tablecaption{Gaussian Models}
\tablewidth{0pt}
\tablehead{
\colhead{component} & \colhead{s} &
 \colhead{r} & \colhead{$\theta$} &
 \colhead{a} & \colhead{b/a} & \colhead{$\phi$}& $\chi_{\nu}^2$\\
            &(mJy) &(mas)&(deg) &(mas) &   &(deg)&
           }
\startdata
         \underline{$\nu$ = 22 GHz} &  &     &  &  &  &\\
           D  & 284$\pm$28 & 0    & 0    & 0.09 & 0.52 & -58&1.00 \\
           C3 & 35$\pm$10 & 0.39$\pm$0.04 & -96$\pm5$ & 0.48 & 0.42 & 70 \\
           C2 & 60$\pm$15 & 1.30$\pm$0.04 & -109$\pm3$ & 0.74 & 0.56 & -26 \\
           C2B& 34$\pm$15 & 2.06$\pm$0.10 & -108$\pm4$ & 0.97 & 1.0  & 0\\
           C1 & 35$\pm$20 & 3.07$\pm$1.00 & -140$\pm6$ & 2.25 & 1.0  & 0 \\
\hline
        \underline{$\nu$ = 15 GHz} &  &     &  &  &  &\\
           D  & 273$\pm$28 & 0    & 0    & 0.22 & 0.61 & 70&1.47 \\
           C3 & 17$\pm$13 & 0.57$\pm$0.30 & -101$\pm5$ & 0.36 & 0.00 & -44&\\
           C2 & 80$\pm$10 & 1.13$\pm$0.06 & -111$\pm3$ & 0.71 & 0.71 & -11&\\
           C2B& 82$\pm$12 & 1.936$\pm$0.14 & -109$\pm4$ & 1.36 & 1.0  & 0 &\\
           C1 & 34$\pm$12 & 3.61$\pm$0.33 & -142$\pm6$ & 1.88 & 1.0  & 0&\\
    \hline
         \underline{$\nu$ = 5 GHz} &  &     &  &   &  &\\
           D  & 342$\pm$50 & 0    & 0    & 0.35 & 1.0  & 0& 2.29\\
           C2 & 367$\pm$50 & 0.95$\pm$0.10 & -120$\pm6$ & 0.70 & 1.0  & 0&\\
           C2B& 173$\pm$30 & 2.33$\pm$0.17 & -116$\pm6$ & 1.30 & 1.0  & 0&\\
           C1 & 088$\pm$30 & 3.51$\pm$1.00 & -150$\pm10$ & 2.02 & 1.0  & 0&\\
           B1 & 053$\pm$40 & 3.11$\pm$1.10 & -44$\pm10$  & 3.34 & 1.0  & 0&\\
\enddata

\tablecomments{ Col.(1) Component name; col.(2) Flux density in Jy
               Col.(3) Distance from core in mas; Col.(4) Position
               angle with respect to the core; Col.(5) Major axis of the fitted
               component in mas; Col.(6) Axis ratio of the fitted component;
               Col.(7) Position angle of the component's major axis;
               Col.(8) Reduced $\chi^2$ of the fit.}

\end{deluxetable}

\begin{deluxetable}{ccccccc}
\tabletypesize{\scriptsize}
\tablecaption{Doppler boosting factors}
\tablewidth{0pt}
\tablehead{
         & 22.22  &15.33 & 8.55 & 5.0 & 2.3 & 1.6  \\
         & (GHz) & (GHz) &(GHz) &(GHz) &(GHz) &(GHz) \\
            VLBI data & this work &this work& FC &this work&FC&PW\\
         }
\startdata
           $\delta_{sp}$  & 9.8 & 3.0 & 9.2 & 5.5 & 6.6 & 9.1  \\
           $\delta_{eq}$  & 3.5 & 0.7 & 4.5 & 2.5 & 3.3 & 5.7 \\
           $\delta_{jet}$ & 5.0 & 2.3 & 6.0 & 4.5 & 5.3 & 7.1\\

\tablecomments{ The reference of VLBI core data:
                  FC   \citep{ Fey97};
                  PW \citep{Pol95}}


\enddata
\end{deluxetable}

\clearpage
\begin{figure}
 \plottwo{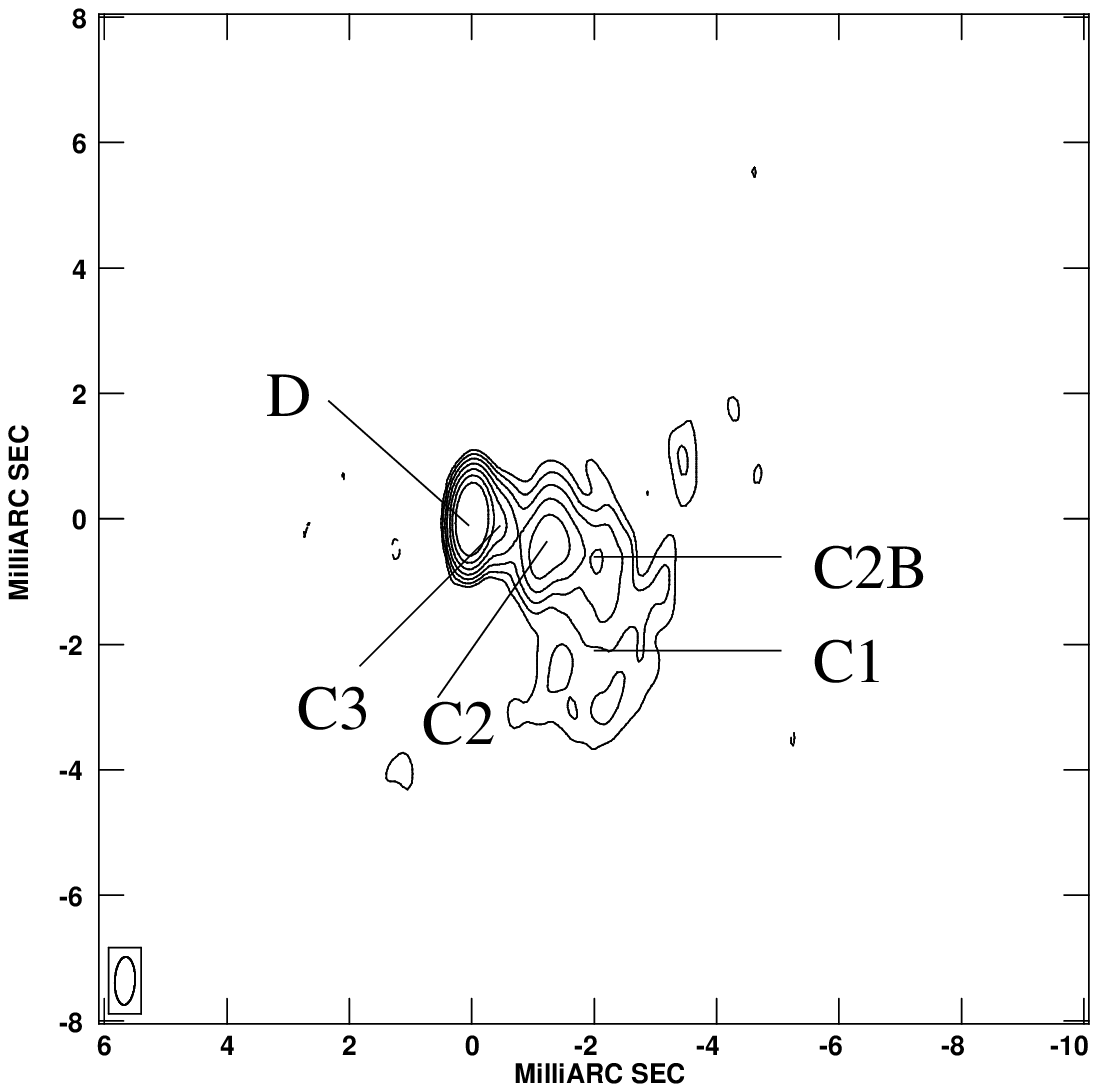}{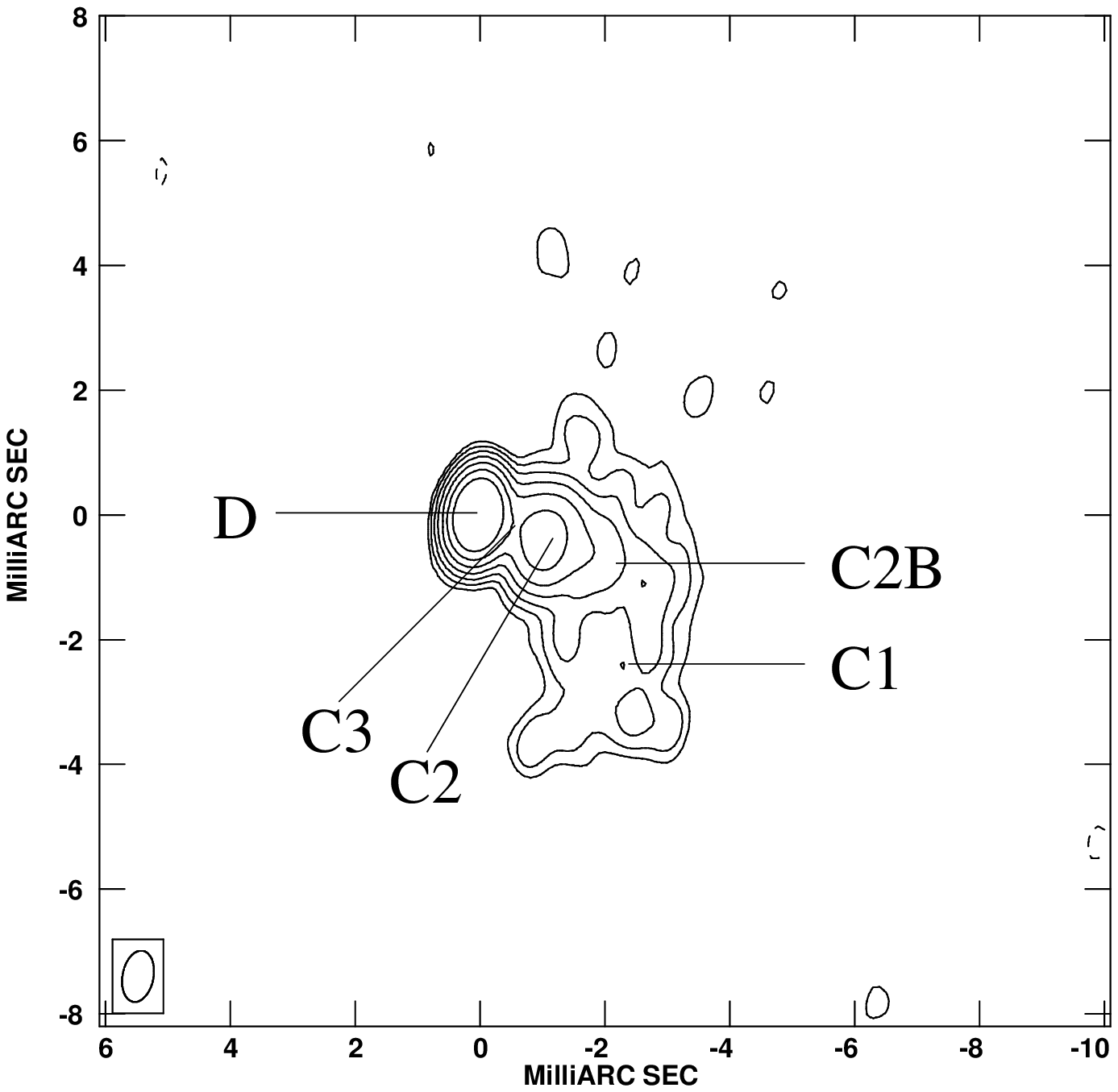}
 \caption{
 Left: the naturally weighted image
 of J1625+4134 at 22~GHz, the restoring beam is $0.77 \times 0.33$~mas
 at PA=$- 2.\degr8$, the contour levels are ($-$1, 1, 2, 4, 8, 16, 32, 64)
$\times 0.9$~mJy/beam, the peak flux density is 0.282~Jy/beam.
Right: the naturally weighted image
 of J1625+4134 at 15~GHz, the restoring beam is
$0.83 \times 0.50$~mas at PA=$- 11.\degr$, the contour levels are
($-$1, 1, 2, 4, 8, 16, 32, 64) $\times 1$~mJy/beam, the peak flux
density is 0.246~Jy/beam.
\label{fig1}}
\end{figure}

\clearpage
\begin{figure}
\plotone{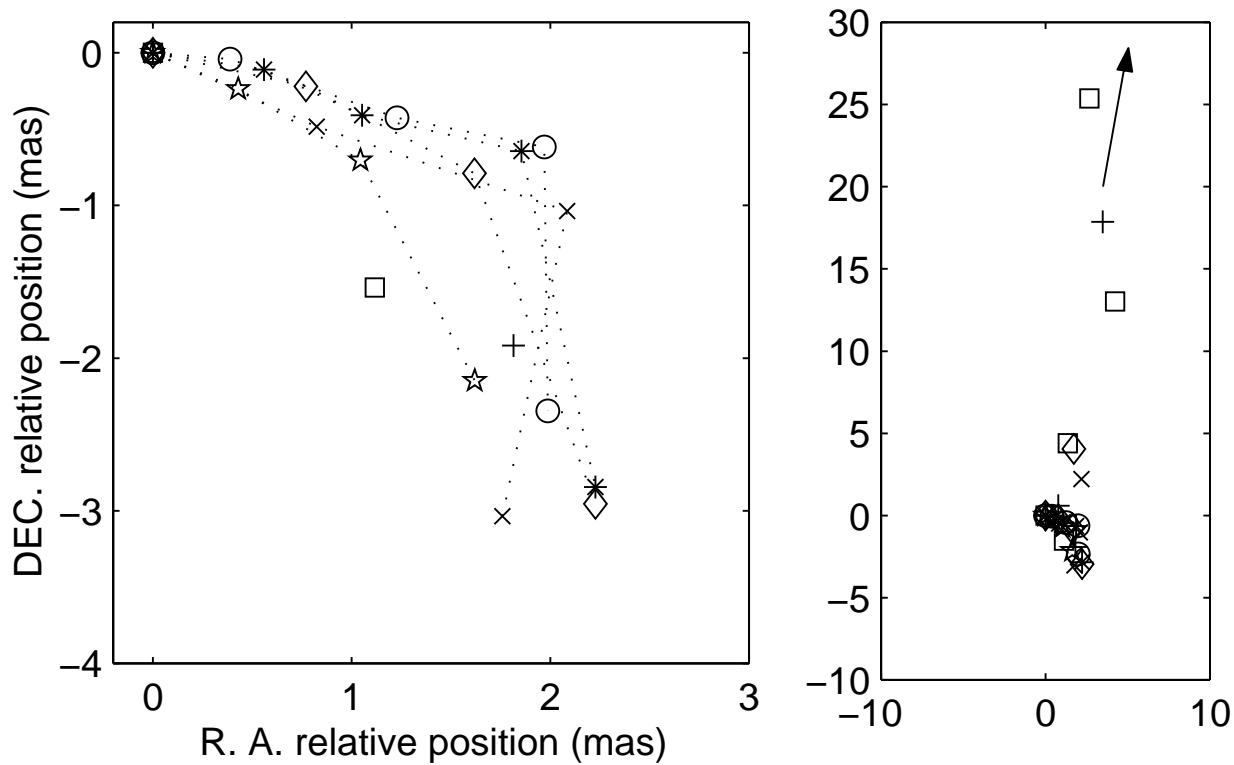} \caption{The position of all jet components
observed in J1625+4134 at various frequencies: 22 GHz (circles,
epoch 2000 Mar 1, this work), 15 GHz (asterisks, epoch 1999 Jan
2, this work), 8.5 GHz (diamonds, epoch 1995 Oct 12, Fey $\&$
Charlot 1997), 5 GHz (crosses, epoch 1996 Jun 6, this work and
Fomalont et al. 2000), VSOP 5 GHz (stars, epoch 1998 Feb 7,
Lister et al. 2001a), 2.3 GHz (squares, epoch 1995 Oct 12, Fey
$\&$ Charlot 1997) and 1.6 GHz (plus, epoch 1990 Sep 21,
Polatidis et al. 1995). The arrow indicates the position angle of
the emission at the arcsecond scale in the VLA images (O'Dea et
al, 1988, Perley 1982). \label{fig2}}
\end{figure}

\clearpage
\begin{figure}
\plotone{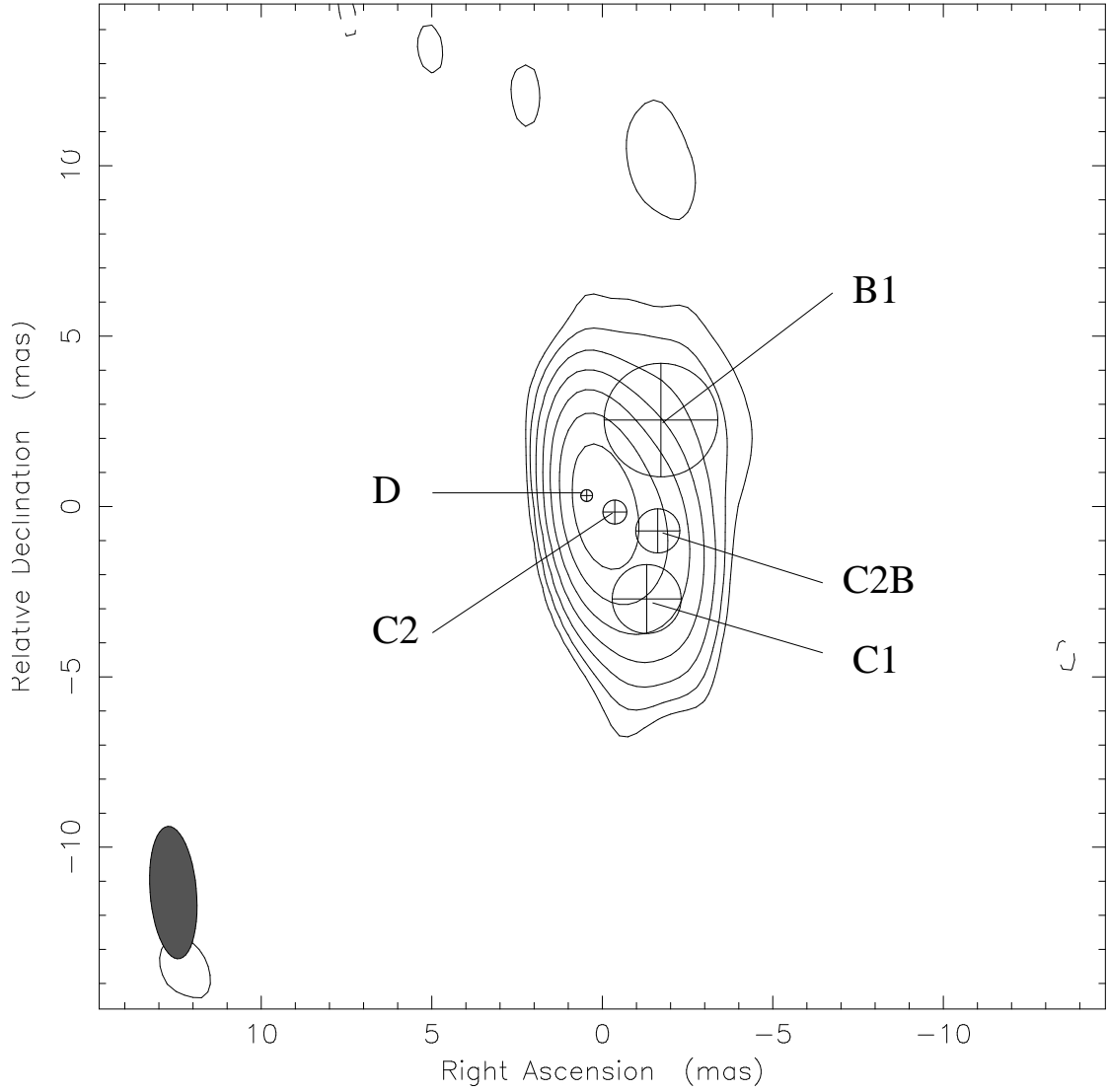} \caption{ The model fitting image of J1625+4134
based on the 5~GHz VLBApls data set (Fomalont et al. 2000). The
restoring beam is $3.9 \times 1.36$~mas at PA=$4.7\degr$, the
contour leves are ($-$1, 1, 2, 4, 8, 16, 32, 64) $\times
5$~mJy/beam, the peak flux density is 0.556~Jy/beam.\label{fig3}}

\end{figure}

\clearpage
\begin{figure}
\plotone{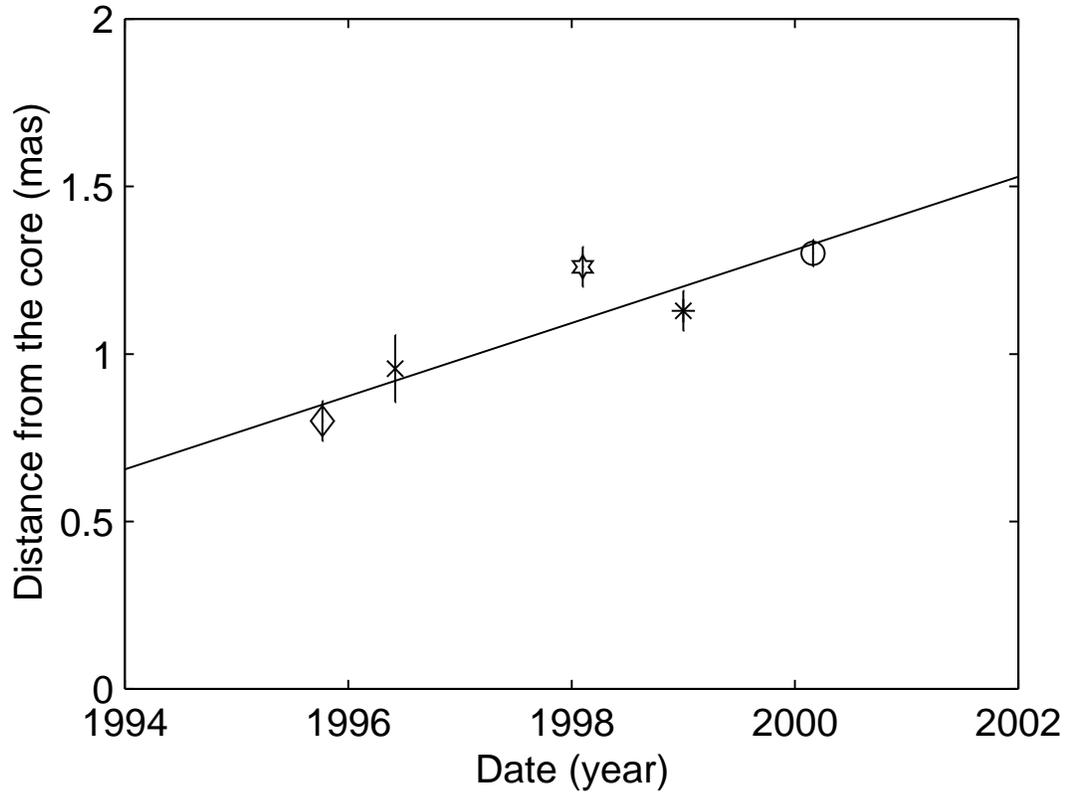} \caption{Apparent proper motion of the jet
component C2 in J1625+4134: 22 GHz (circle, epoch 2000 Mar 1, this
work), 15 GHz (asterisk, epoch 1999 Jan 2, this work), 8~GHz
(diamond, epoch 1995 Oct 12, Fey $\&$ Charlot 1997), 5~GHz
(cross, epoch 1996 Jun 6, this work and Fomalont et al. 2000) and
VSOP 5~GHz (star, epoch 1998 Feb 7, Lister et al. 2001a). The
position errors of the C2 component at 15~GHz, 22~GHz and 5~GHz
are in the table 1, and for the errors of the C2 component at
8~GHz and VSOP 5~GHz, a value 0.06 mas is assumed.\label{fig4}}

\end{figure}

\clearpage

\begin{figure}
\plotone{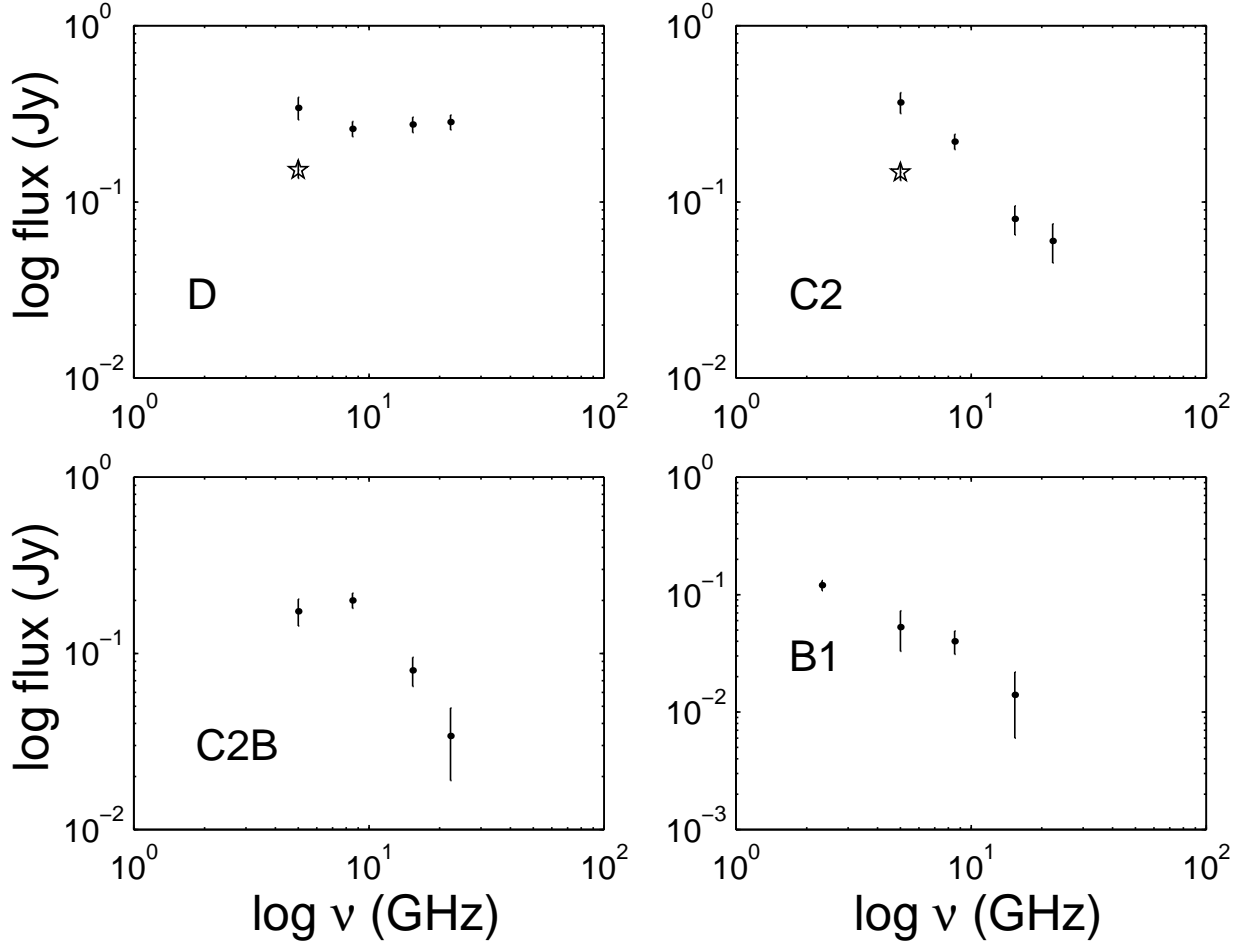} \caption{Spectra of the jet components in
J1625+4134. The stars present the flux density values of the
components from VSOP 5 GHz observation. The error estimates of the
flux density of the components at 15~GHz, 22~GHz and 5~GHz are in
the table 1, and a 10 percent of the flux density for the
components at 8~GHz and VSOP 5~GHz are assumed as the estimates
of the errors.
 \label{fig5}}
\end{figure}

\clearpage

\begin{figure}
\plotone{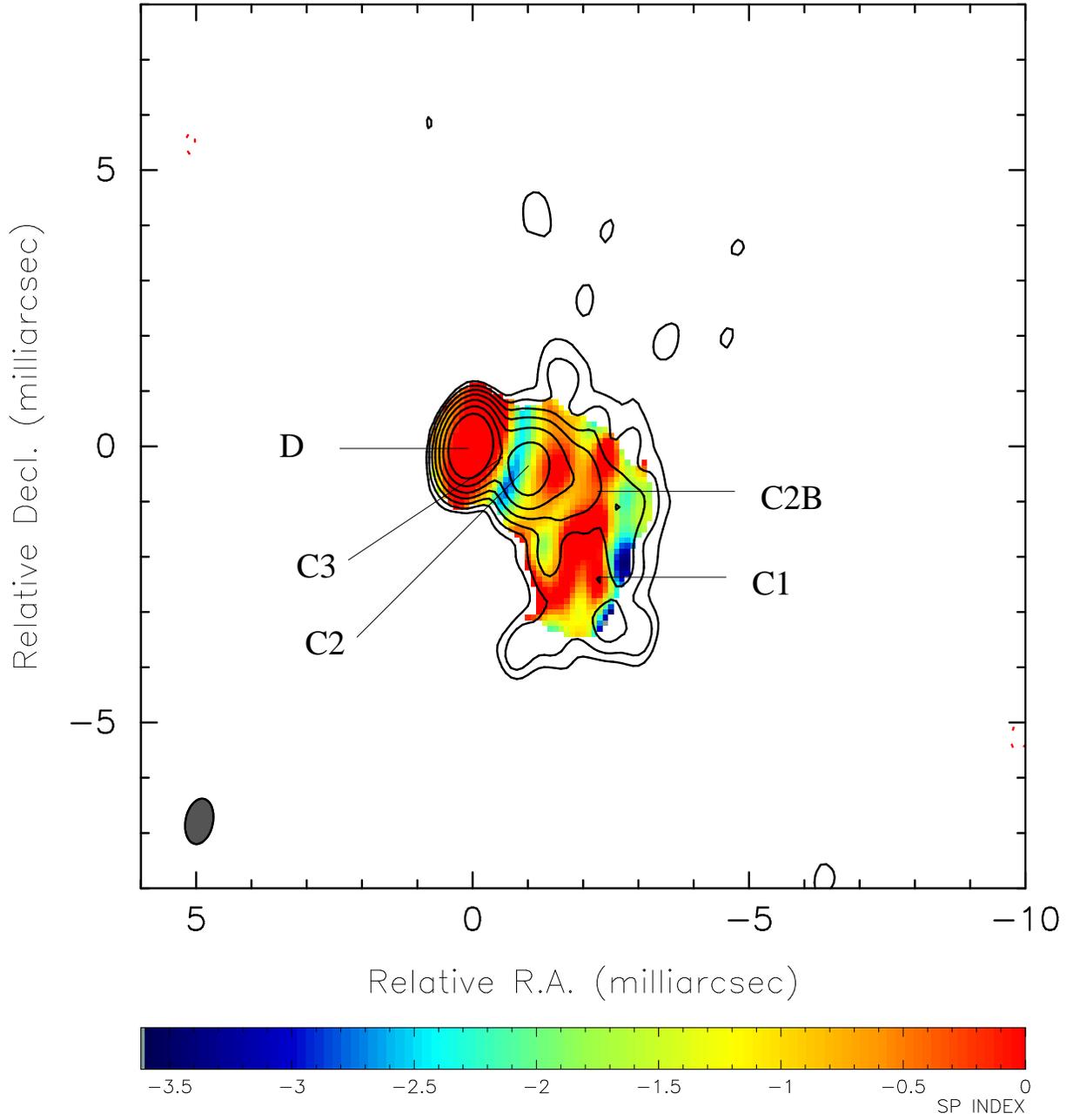} \caption{The non-simultaneous two-frequency
spectral index distribution between 15 and 22~GHz. The wedge at
the bottom shows the spectral index scale. The contours are from
the 15~GHz image (Fig.~1). \label{fig6}}
\end{figure}

\clearpage

\begin{figure}
\plotone{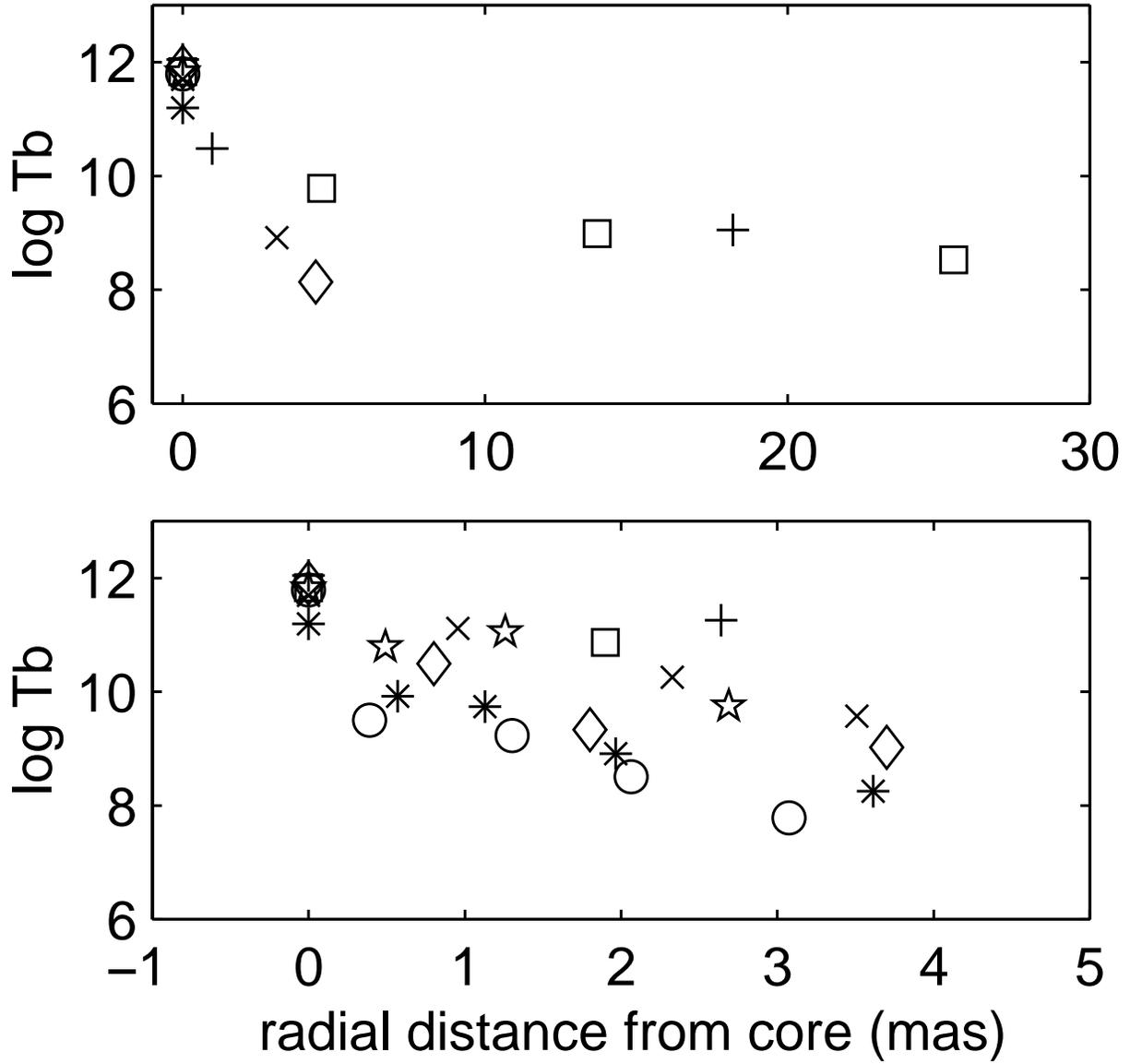} \caption{The source rest-frame brightness
temperature  distribution of the VLBI components in the jet of
J1625+4134. The upper panel shows the northern jet, the lower
panel shows the southwest (inner). The symbols are the same as in
Fig.~2.\label{fig7}}

\end{figure}


\begin{thebibliography}{}
\bibitem[Biermann et al. 1987]{Bie87}Biermann, P. I., K\"uhr, H., Snyder, W. A., \& Zensus, J. A., 1987, \aap, 185, 9
\bibitem[Brinkmann et al. 2000]{Brink2000}Brinkmann, W., Laurent-Muehleisen, S. A.
         Voges, W., Siebert, J.,et al., 2000, \aap, 356, 445
\bibitem[Cao 2000]{Cao2000}Cao Xinwu, 2000, \aap, 355, 44
\bibitem[Conway \& Murphy 1993]{Conway93}Conway, J. E. \& Murphy, D. W., 1993, \apj, 411, 89
\bibitem[Fey \& Charlot 1997]{Fey97}Fey, A. L. \& Charlot, P., 1997, \apjs, 111, 95
\bibitem[Frey et al. 2002]{Frey02}Frey, S., et al., 2002, in preparation
\bibitem[Fomalont et al. 2000]{Foma2000}Fomalont, E. B., Frey, S., Paragi, Z.,  et al., 2000, \apjs, 131, 95
\bibitem[1988]{Foma88}Fomalont, E. B., 1988, in Synthesis Imaging in Radio Astronomy, edited by R. A. Perley, F. R. Schwab
         and A. H. Bridle, 213
\bibitem[1993]{Ghis93}Ghisellini, G, Padovani, P, Celloti, A., 1993, \apj, 407, 65
\bibitem[G\'omez et al. 1990]{Gome99}G\'omez, J., Marscher, A. P., Alberdi, A., and Gabuzda, D. C., 1999, \apj, 519, 642
\bibitem[G\"uijosa and Daly 1996]{Gui96}G\"uijosa, A. \& Daly, R. A., 1996, \apj, 461, 600
\bibitem[Gurvits et al. 2002]{Gur2002}Gurvits, L.I., Kellermann, K.I., Fomalont, E.B. and  Zhang, H.Y., 2002, in preparation
\bibitem[Hewitt \& Burbidge 1993]{Hewitt93}Hewitt, A., \& Burbidge, G., 1993, \apjs, 87, 451
\bibitem[Hong et al. 1998]{Hong98}Hong, X. Y., Jiang, D. R. and Shen, Z. Q., 1998, \aap, 330, L45
\bibitem[1998]{Jiang98}Jiang, D. R., Cao, X. and Hong, X., 1998, \apj, 494, 139
\bibitem[K\"onigl 1981]{Koni81}K\"onigl, A., 1981, \apj, 243, 700
\bibitem[Krichbaum et al. 1999]{Kric99}Krichbaum, T. P. et al.,1999, in 2nd Millimeter-VLBI Science Workshop,
          held at IRAM Granada, Spain, 27-29, eds. A. Greve and T. P. Krichbaum, 5
\bibitem[Lebofsky et al. 1983]{Lebo83}Lebofsky, M. J., Rieke, G. H. and Walsh, D., 1983 \mnras, 203, 727
\bibitem[Lister et al. 2001a]{Lis2001a}Lister, M. L., Tingay, S. J., Preston, R. A., 2001a, \apj, 554, 948
\bibitem[Lister et al. 2001b]{Lis2001b}Lister, M. L., Tingay, S. J., Preston, R. A., 2001b, \apj, 554, 964
\bibitem[Lobanov 1998]{Loba98}Lobanov, A. P., 1998, \aap, 330, 79
\bibitem[Marcher 1987]{Mar87}Marcher, A. P., 1987, in Superluminal Radio Sources, eds by Zensus J. A. and
          Pearson, T. J., Cambridge University Press, New York, 280
\bibitem[O'Dea et al. 1988]{ODea88}O'Dea, C. P., Barvainis, R., Challis, P. M., 1988, \aj, 96, 435
\bibitem[Pearson \& Readhead 1988]{Pear88}Pearson, T. J. and Readhead, A. C. S., 1988, \apj, 328, 114
\bibitem[Perley 1982]{Per82}Perley, R. A., 1982, \aj, 87, 859
\bibitem[Polatidis et al. 1995]{Pol95}Polatidis, A. G., Wilkinson, P. N. et al., 1995, \apjs, 98, 1
\bibitem[1994]{Read94}Readhead, A. C. S., 1994, \apj, 426, 51
\bibitem[Sambruna et al. 1996]{Sam96}Sambruna, R. M., Maraschi, L., Urry, C. M., 1996, \apj, 463, 444
\bibitem[Shen et al. 1997]{Shen97}Shen, Z.-Q., Wan, T. S., Moran, J. M., Jauncey, D. L., et al., 1997, \aj, 114, 1999
\bibitem[Shepherd et al. 1994]{Shep94}Shepherd, M. C., Pearson, T. J., \& Taylor, G. B., 1994, \baas, 26, 987


\end{thebibliography}
\end{document}